\documentclass[sigconf]{acmart}
\author[J. Xun*, S. Zhang*, Z. Zhao, J. Zhu, Q. Zhang, J. Li, X. He, X. He, T. Chua, F. Wu]{
    Jiahao Xun$^{1*}$, Shengyu Zhang$^{1*}$, Zhou Zhao$^{1\dagger}$, Jieming Zhu$^{2}$, Qi Zhang$^{2}$, Jingjie Li$^{2}$, Xiuqiang He$^{2}$, Xiaofei He$^{1}$, Tat-Seng Chua$^{3}$, Fei Wu$^{1}$
}
\affiliation{
  $^1$ Zhejiang University
  \country{China}
}
\affiliation{
  $^2$ Huawei Noah's Ark Lab
  \country{China}
}
\affiliation{
  $^3$ National university of Singapore
  \country{Singapore}
}
\email{
  {jhxun, sy_zhang, zhaozhou, wufei}@zju.edu.cn
}

\email{
  {jamie.zhu, zhangqi193, lijingjie1, hexiuqiang1}@huawei.com
}
\email{
  xiaofeihe@cad.zju.edu.cn
}
\email{dcscts@nus.edu.sg}

\renewcommand{\authors}{Jiahao Xun, Shengyu Zhang, Zhou Zhao, Jieming Zhu, Qi Zhang, Jingjie Li, Xiuqiang He, Xiaofei He, Tat-Seng Chua, Fei Wu}

\AtBeginDocument{%
  \providecommand\BibTeX{{%
    \normalfont B\kern-0.5em{\scshape i\kern-0.25em b}\kern-0.8em\TeX}}}

\copyrightyear{2021}
\acmYear{2021}
\setcopyright{acmcopyright}\acmConference[MM '21]{Proceedings of the 29th ACM International Conference on Multimedia}{October 20--24, 2021}{Virtual Event, China}
\acmBooktitle{Proceedings of the 29th ACM International Conference on Multimedia (MM '21), October 20--24, 2021, Virtual Event, China}
\acmPrice{15.00}
\acmDOI{10.1145/3474085.3475514}
\acmISBN{978-1-4503-8651-7/21/10}

\usepackage{multirow}
\usepackage{footmisc}
\usepackage{subcaption,siunitx,booktabs}
\usepackage{caption}
\usepackage{bbm}

\usepackage{multirow}
\usepackage{balance}
\usepackage{enumitem}

\begin{document}
\fancyhead{}

\title{ Why Do We Click: Visual Impression-aware News Recommendation }


\newcommand{\etal}{\textit{et al}.}
\newcommand{\ie}{\textit{i}.\textit{e}.}
\newcommand{\eg}{\textit{e}.\textit{g}.}
\newcommand{\vpara}[1]{\vspace{0.05in}\noindent\textbf{#1 }}
\newcommand\jiahao[1]{\textcolor{red}{[Jiahao: #1]}}



\begin{abstract}

There is a soaring interest in the news recommendation research scenario due to the information overload. To accurately capture users' interests, we propose to model multi-modal features, 
in addition to the news titles that are widely used in existing works, for news recommendation. 
Besides, existing research pays little attention to the click decision-making process in designing multi-modal modeling modules. In this work, inspired by the fact that users make their click decisions mostly based on the visual impression\footnote{In this paper, we refer to \textit{visual impression} as the region of the news displayed on the user interface of a news application, which delivers both content and layout information to users.}
they perceive when browsing news, we 
propose to capture such visual impression information with visual-semantic modeling for news recommendation. Specifically, we devise the local impression modeling module to simultaneously attend to decomposed details in the impression when understanding the semantic meaning of news title, which could explicitly get close to the process of users reading news. In addition, we inspect the impression from a global view and take structural information, such as the arrangement of different fields and spatial position of different words on the impression, into the modeling of multiple modalities. To accommodate the research of visual impression-aware news recommendation, we extend the text-dominated news recommendation dataset MIND by 
adding snapshot impression images
and will release it to nourish the research field. Extensive comparisons with the state-of-the-art news recommenders along with the in-depth analyses demonstrate the effectiveness of the proposed method and the promising capability of modeling visual impressions for the content-based recommenders.
\renewcommand{\thefootnote}{\fnsymbol{footnote}}
\footnotetext[1]{These authors contributed equally to this work.}
\footnotetext[2]{Corresponding Author.}
\end{abstract}





\maketitle

\section{Introduction}

Nowadays, online content sharing platforms have changed the way of people reading news in a mobile and digital manner. News production sources have been extremely enlarged on such platforms, such as Microsoft News\footnote{https://www.msn.com/en-us/news} and Google News\footnote{https://news.google.com}, that users 
could suffer from information overload due to the overwhelming amount of news.
To mitigate information overload and improve user experiences, personalized news recommender systems are devised to make it easy for users to find the news of their interests. 
The challenging and open-ended nature of news recommendation lends itself to diverse advances in the literature
\cite{Okura_Tagami_Ono_Tajima_2017,Wang_Wu_Liu_Xie_2020,Wang_Zhang_Xie_Guo_2018,Wu_Wu_An_Huang_Huang_Xie_2019a,Wu_Wu_An_Huang_Huang_Xie_2019b,Wu_Wu_Ge_Qi_Huang_Xie_2019,Zhu_Zhou_Song_Tan_Guo_2019, yao2021device}. 
\begin{figure}[!t] \begin{center}
    \includegraphics[width=0.95\columnwidth]{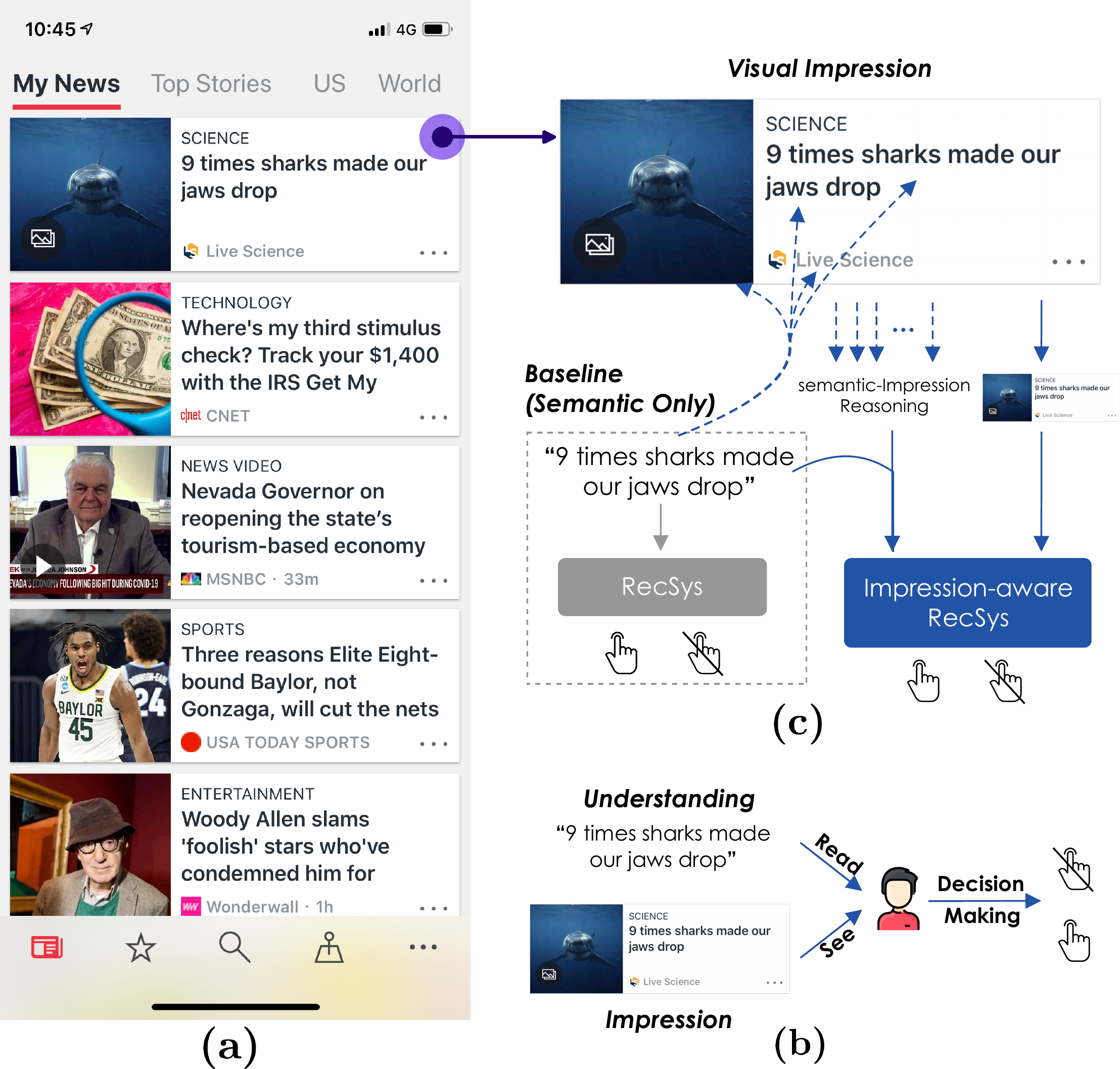}
    \caption{
    An illustration of impression-aware news recommendation. (a) The interface that users are browsing. (b) Before making click decisions, users typically have the semantic understanding of news title and visual impression in mind. (c) Impression-aware recommendation takes the fine-grained visual cues and the global structures into account. 
    	}
\label{fig:firstpage}
\vspace{-1cm}
\end{center} \end{figure}
Recently, Okura \etal, \cite{Okura_Tagami_Ono_Tajima_2017} learn to represent historically interacted news for a user via a denoising autoencoder and RNNs 
in the recommender system of Yahoo! JAPAN\footnote{https://www.yahoo.co.jp}. Wang \etal, \cite{Wang_Wu_Liu_Xie_2020} learn to obtain multi-level user representations with stacked dilated convolutions. Despite significant progress made with these advances, they solely use the 
textual contents of news titles to represent users' interests and ignore the digital news's multi-modal nature. As shown in Figure \ref{fig:firstpage}, online news might contain a variety of modalities or fields, \ie, title,  body, video, soundtrack, image, and category. Thus, we derive inspirations from many other domains \cite{li2021adaptive, jin2021hierarchical, zhang2021causerec, lu2021future,zhang2020does,jin2019multi,zhang2020comprehensive, li2019walking,li2020unsupervised,tian2021analysis,zhang2020counterfactual,zhang2021learning} and propose to incorporate multi-modal information for an in-depth understanding of users' preferences on news.

Recently, advances in other domains and applications have demonstrated great successes of multi-modal recommender systems \cite{Arapakis_Moshfeghi_Joho_Ren_Hannah_Jose_2009,Chelliah_Biswas_Dhakad_2019,Kuo_Shan_Lee_2013,Wang_Zhao_Nie_Gao_Nie_Zha_Chua_2015,Wei_Wang_Nie_He_Chua_2020,Wei_Wang_Nie_He_Hong_Chua_2019,Yu_Shen_Zhang_Zeng_Jin_2019,Zhao_Yang_Lu_Weninger_Cai_He_Zhuang_2018}. For example, Wei \etal, \cite{Wei_Wang_Nie_He_Hong_Chua_2019} propose to model individual user-item interactions for each modality and use graph convolutional networks \cite{kipf2016semi} to learn modality-specific representations. Following this work, Wei \etal, \cite{Wei_Wang_Nie_He_Chua_2020} propose to refine the user-item graph connections for each modality and thus leverage modality-specific network structures, which also helps mitigate the implicit feedbacks. Zhao \etal, \cite{Zhao_Yang_Lu_Weninger_Cai_He_Zhuang_2018} propose to learn multi-modal heterogeneous network representation and incorporate user profiles, social relationships, textual description, and video posters for video recommendation. Despite their successes on real-world datasets, we argue that these methods have two major deficiencies. Firstly, they typically introduce all available modalities without evaluating and explaining which modalities are essential for click-through-rate (CTR) prediction. However, introducing more features would not necessarily mean being more effective since such features might lead to expensive computation, more over-fitting, and even more noise. Secondly, most of them tend to leverage modalities with generic architectures with less recommendation-specific or application-specific designs. 

Towards this end, we propose to investigate which modalities we should incorporate for news recommendation and design fusion modules with highly application-specific insights. We aim to answer the question, "why do users click" and start with the perspective that the user's click decision is mostly based on his/her inherent interest and the visual impression delivered by the news.
The visual impression can be the visual-semantic information he/she perceives when browsing the news application. In this paper, we treat the visual region of the news displayed on the user interface of news applications as the visual impression (as shown in Figure \ref{fig:firstpage}). 
Therefore, our work aims to model such multi-modal visual impression information to improve
the click-through-rate prediction. We contend that other modalities or fields such as news body and soundtrack are inaccessible before users clicking the news. Recommender systems might draw false conclusions when spuriously connecting these modalities to users' interests. Furthermore, we leverage the layout information such as relative positions, relative sizes, and styles as guidance for multi-modal fusion, \ie, a news recommendation-specific design. To be specific, we devise the \textbf{\textit{IM}}pression-aware multi-modal news \textbf{\textit{Rec}}ommendation framework, denoted as IMRec. IMRec comprises two key components: (1) The global impression module that not only fuses the multi-modal content features under the guidance of news layout but also enhances the global item representations. (2) The local impression module that models the correlation of 
each title word and other impression units, such as visual title words and visual images. In this way, our model bridges the gap between semantic understanding and visual impressions for each news in a fine-grained manner. 

To the best of our knowledge, this work is one of the initiatives to investigate impression-aware recommendation, and there is currently no news recommendation dataset suitable for this research. To this end, we construct a large-scale impression-aware news recommendation dataset, IM-MIND, by adding the snapshot impression images into a text-dominated
benchmark, MIND \cite{Wu_Qiao_Chen_Wu_Qi_Lian_Liu_Xie_Gao_Wu_et_2020}. We conduct in-depth experimental analyses along on 
both quantitative and qualitative results, which have demonstrated the effectiveness and necessity of modeling visual impressions for news recommendation. The highlights of this work are summarized
as follows:

\begin{itemize}
	\item We discuss why users click a news article at an intuitive level and propose to investigate impression-aware news recommendation, which guides the modality selection and model design better for click-through-rate prediction.
	\item We propose the IMRec framework that comprehensively exploits the visual impression features in a global-local manner and bridges the gap between the semantic meaning of news title and the visual impression users perceive before clicking.
	\item We contribute the visual images of news impressions to MIND dataset to facilitate this line of research and demonstrate the effectiveness of IMRec framework with extensive experiments.
\end{itemize}

Noteworthy, for ease of modeling, we currently model the visual impression of each news independently because the surrounding news articles can only be obtained after they are finally ranked and displayed to the users. During training and inference, we could simulate the visual impression of each news through the software of UI interface.

\section{Related Works}

\subsection{News Recommendation}

In recent years, the explosively growing amount of digital news calls for effective news recommender systems which enable personalized news suggestions. Both natural language processing and data mining research fields \cite{Phelan_McCarthy_Bennett_Smyth_2011,Zheng_Zhang_Zheng_Xiang_Yuan_Xie_Li_2018,Wu_Wu_An_Huang_Xie_2019} have witnessed deep learning based models' successes in extracting semantic content features and mining user preferences accordingly \cite{Zhu_Zhou_Song_Tan_Guo_2019,Wu_Wu_Wang_Huang_Xie_2020,Wu_Wu_An_Qi_Huang_Huang_Xie_2019,Wu_Wu_An_Huang_Huang_Xie_2019b,Wu_Wu_An_Huang_Huang_Xie_2019a,Wang_Zhang_Xie_Guo_2018,Wang_Wu_Liu_Xie_2020,Okura_Tagami_Ono_Tajima_2017,Hu_Xu_Li_Yang_Shi_Duan_Xie_Zhou_2020,An_Wu_Wu_Zhang_Liu_Xie_2019}. Diverse models concerning RNNs \cite{Okura_Tagami_Ono_Tajima_2017}, attention mechanisms \cite{Wu_Wu_An_Huang_Huang_Xie_2019b,Wu_Wu_An_Huang_Huang_Xie_2019a,Zhu_Zhou_Song_Tan_Guo_2019}, dilated convolution \cite{Wang_Wu_Liu_Xie_2020}, graph neural networks \cite{Hu_Xu_Li_Yang_Shi_Duan_Xie_Zhou_2020}, and knowledge distillation \cite{Wang_Zhang_Xie_Guo_2018} are explored. Typically, Wu \etal, \cite{Wu_Wu_Ge_Qi_Huang_Xie_2019} leverage both self-attention mechanisms \cite{Vaswani_Shazeer_Parmar_Uszkoreit_Jones_Gomez_Kaiser_Polosukhin_2017} and additive attention \cite{Bahdanau_Cho_Bengio_2015} to represent words with one news and multiple news with the user's historical interactions. FIM is a state-of-the-art recommendation model proposed by Wang \etal, \cite{Wang_Wu_Liu_Xie_2020} that captures fine-grained interest matching signals using dilated convolutions. However, most of these works solely model the news title and disregard other modalities that might highly contribute to user's click behavior, such as the news cover image. Towards this end, we propose to incorporate necessary modalities and design news recommendation-specific architectures.

\subsection{Multi-modal Recommendation}

Online content sharing platforms are becoming rich in modalities due to the rapidly developing network communication technologies. Therefore, as a nascent research field, multi-modal recommendation attracts increasing attention \cite{Arapakis_Moshfeghi_Joho_Ren_Hannah_Jose_2009,Yu_Shen_Zhang_Zeng_Jin_2019, zhang2020devlbert} recently with applications in various domains such as music recommendation \cite{Kuo_Shan_Lee_2013}, location recommendation \cite{Wang_Zhao_Nie_Gao_Nie_Zha_Chua_2015}, movie recommendation \cite{Zhao_Yang_Lu_Weninger_Cai_He_Zhuang_2018}, micro-video recommendation \cite{Wei_Wang_Nie_He_Hong_Chua_2019,Wei_Wang_Nie_He_Chua_2020}, and fashion recommendation \cite{Chelliah_Biswas_Dhakad_2019}. Noteworthy, MMGCN \cite{Wei_Wang_Nie_He_Hong_Chua_2019} and GRCN \cite{Wei_Wang_Nie_He_Chua_2020} construct a user-item bipartite graph and conduct information propagation and embedding learning for each modality. GRCN differs from MMGCN by refining each graph's connections and denoising implicit feedback in the fine-grained modality level. Despite great successes, we argue that most of them model multiple modalities without domain knowledge guidance. To be specific, most architectures disregard a fundamental question, "why do users click" and fail to model the impression which can be necessary for users' decision making. In this paper, we propose the impression-aware recommendation framework designed especially for news recommendation by explicitly modeling users' click decision-making processes.

\section{Methods}

\subsection{Problem Formulation}

Following the common practice in modern news recommender systems \cite{Wu_Wu_An_Huang_Huang_Xie_2019b,Wu_Wu_Ge_Qi_Huang_Xie_2019}, we formulate news recommendation as a sequential recommendation problem and specifically focus on the news click-through-rate (CTR) prediction. We use $u$ to denote one user and $\mathcal{I}_u = \{ e_t \}_{t=1,\dots,|\mathcal{I}_u|}$ to denote the sequence of news historically clicked by user $u$ on an online news platform. $\mathcal{I}_u$ is ordered by the click time beforehand. News 
CTR prediction aims to predict whether the user $u$ will click a candidate news $c$ with the binary label denoted as $y_i \in \{ 0, 1 \}$. A deep learning based news recommendation model $f(\cdot)$ takes a pair of user $u$ and candidate news $c$ as input and predicts a probability $\hat y = f(u,c)$ indicating how likely the click will happen. During testing and serving, candidate news will be ranked by the probabilities and displayed on the news platform with positions consistent with the ranks.

\begin{figure}[!t] \begin{center}
    \includegraphics[width=\columnwidth]{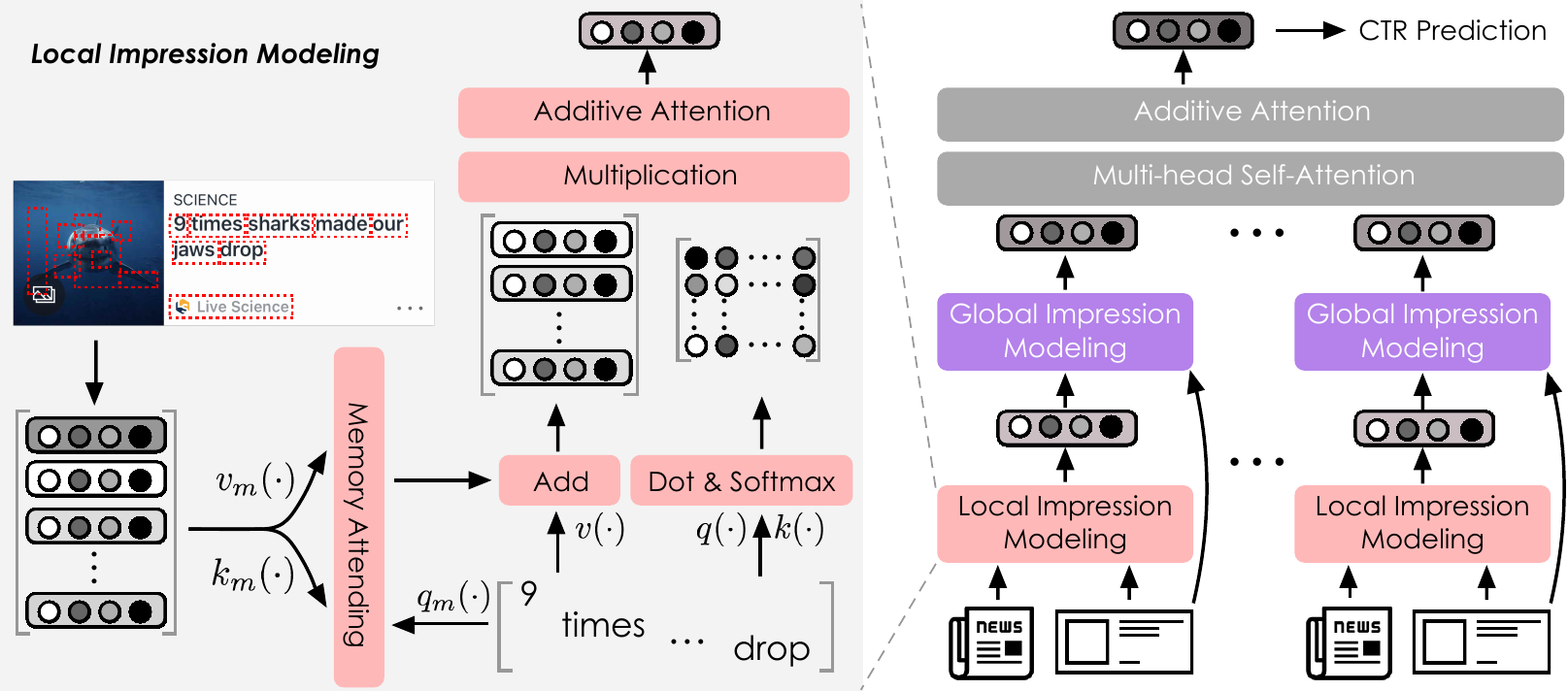}
    \caption{
    	Schematic illustration of impression-aware news recommendation framework applied to NRMS. We represent each news using \textit{local impression modeling}, which explicitly captures multi-modal visual cues within the visual impression and accordingly enhances the semantic understanding of news title, and \textit{global impression modeling}, that models the visual impression as whole by further taking the arrangement of different fields and relative position of title words into consideration.
	}
\label{fig:schema}
\vspace{-0.5cm}
\end{center} \end{figure} 


\subsection{Impression-aware Recommendation}

To explicitly 
model users' click decision-making processes (depicted in Figure \ref{fig:firstpage}) and bridge the gap between semantic understanding and visual impression of news, we devise the impression-aware recommendation framework, denoted as IMRec. As depicted in Figure \ref{fig:schema}, NRMS with IMRec framework (denoted as \textbf{NRMS-IM}) incorporates the local details of visual impression into the semantic understanding of news titles, which is inspired by the users' browsing processes that users not only read the meaning of titles but 
meanwhile 
receive many impression details such as the visual appearance of words and regions in the news cover image. We denote such a process as \textbf{local impression modeling}. Moreover, once users have captured all the details, they might attempt to construct a holistic recognition of the news, in which we incorporate the fused representation of all modalities to enhance the details further. We denote the holistic impression modeling as \textbf{global impression modeling}, which introduces more structural information from a global view, such as the alignment of different fields and the relative position of words. In the following sections, we will formally describe these two processes based on the sequence model NRMS.

\subsubsection{Local Impression Modeling} Local impression modeling aims to simultaneously capture the local impression details while understanding the semantic meaning of news titles. Towards this end, we encapsulate an \textit{impression decomposition} process that explicitly extracts meaningful cues from the impression image beforehand, and an \textit{impression-semantic reasoning} module that bridges the modality gap and structural gap between impression and semantics.

\vpara{Impression Decomposition.} To ease the modeling of local details in the impression image, we propose to extract meaningful cues in a pre-processing manner, which obtains analogous gains observed in many other domains \cite{Lu_Batra_Parikh_Lee_2019,Anderson_He_Buehler_Teney_Johnson_Gould_Zhang_2018}. However, different from previous works that commonly employ an object detector \cite{Ren_He_Girshick_Sun_2015}, which is designed to process natural scenes, we propose to first divide the impression image into several salient parts and extract cues from the corresponding feature maps. Since the impression image is well structured, we obtain the news title part, the news cover image part, and the news category part with simple edge detection techniques. For the news \textbf{title} part, we view each word region $w^v_i$ as an individual cue that users can potentially attend when understanding the semantic meaning of titles. We denote the vectorial representation of word region $w^v_i$ as $\mathbf{w}^v_i$. For the news \textbf{cover image} part, we view each region vector $\mathbf{v}_i$ in the feature map $\mathbf{V} = \{ \mathbf{v}_i \}_{i=1,\dots,|V|}$ extracted by a pre-trained CNN as a cue. For the news \textbf{category} part, we directly view the whole region $a_i$ with its vectorial representation $\mathbf{a}_i$ as a cue. The details of the pre-processing and used pre-trained architectures can be found in Section \ref{sec:expdetails}. 

To ease modeling, we group all the pre-extracted cues together to construct a impression cue memory $\mathbf{O} = \{\mathbf{o}_i\}_{i=1,\dots,|O|}$. Since the representations of different cues are obtained using the same feature extractor, they naturally belong to the same embedding hypersphere and we treat them equally in the following modeling.

\vpara{Impression-Semantic Reasoning.} Given one news title $e$, comprised of a sequence of words $\{ w_i \}_{1,\dots,|e|}$, we first embedding the sequence into a low-dimensional representation $\mathbf{E} = \{ \mathbf{w}_i \}_{1,\dots,|e|}$. To capture the correlations between impression and semantics, we view the impression cues as the external knowledge and follow the memory network schema \cite{Sukhbaatar_Szlam_Weston_Fergus_2015}:
\begin{align}
	\alpha^v_{ij} = \frac{ \exp{ (q_m( \mathbf{w}_i )^T k_m( \mathbf{o}_j) } ) }{ \sum_k \exp{ (q_m( \mathbf{w}_i )^T k_m( \mathbf{o}_k )) } }, \\
	\mathbf{\hat w}_i = \sum_j \alpha^v_{ij} * v_m( \mathbf{o}_j ) + v(\mathbf{w}_i),
\end{align}
where $q_m(\cdot)$, $k_m(\cdot)$, and $v_m(\cdot)$ denote linear transformations with bias terms. $\alpha^v_{ij}$ denotes the extent to which the user will attend to impression cue $o_j$ when reading the word $w_i$. Summing all attended cues results to the linearly transformed word embedding $v(\mathbf{w}_i)$ in an impression-aware word representation $\mathbf{\hat w}_i$. To further reason on the impression-semantic joint representations, we next leverage the semantic dependencies implied by the self-attended weights:
\begin{align}
	\alpha^v_{ij} = \frac{ \exp{ (q( \mathbf{w}_i )^T k( \mathbf{w}_j) } ) }{ \sum_k \exp{ (q( \mathbf{w}_i )^T k( \mathbf{w}_k )) } },
\end{align}
where $q(\cdot)$ and $k(\cdot)$ are linear transformations and $\alpha^v_{ij}$ denotes the extent to which the model attends to the impression-semantic representation of the $j$th word to enhance the final representation of the $i$th word. The holistic representation of one news is obtained by summing all words with additive attention weights \cite{Bahdanau_Cho_Bengio_2015}:
\begin{align}
	\mathbf{w}^*_i &= \sum_j \alpha^v_{ij} \mathbf{\hat w}_j, \\
	\alpha^a_i &= \frac{\exp q_a( \tanh (k_a (\mathbf{w}^*_i)) )}{\exp \sum_k q_a( \tanh (k_a (\mathbf{w}^*_k)) )}, \\
	\mathbf{e} &= \sum_i \alpha^a_i * \mathbf{w}^*_i
\end{align}
where $k_a$ transforms $\mathbf{w}^*_i$ into a hidden space and $q_a$ computes the attention weights for aggregation.

\subsubsection{Global Impression Modeling.} Local impression modeling captures impression cues separately, which means we disregard the correlations and interactions between different impression cues. A straightforward way is to model them using traditional multi-modal fusion techniques. However, directly employing off-the-shelf techniques might lead to the loss of structural information, such as the location arrangement of different fields and the spatial position of different words. Therefore, instead of fusing different cues separately, we propose to encode the impression image as a whole with pre-trained extractors. Given the global impression embedding $\mathbf{o}^*$, we have:
\begin{align}
	a &= \sigma(g(\mathbf{o}^*, \mathbf{e})), \\
	\mathbf{e}^* &=  a * \mathbf{e} + (1-a) * \mathbf{o}^*.
\end{align}
where $g(\cdot)$ is a linear transformation, $\sigma$ denotes the sigmoid function, and $a$ serves as a gate to control how much information we should let through from $\mathbf{e}$ and $\mathbf{o}^*$, by taking their information into consideration. Such a gate is reasonable in the sense that users might not be equally interested in the impression and the textual semantics, and $a$ indicates a tradeoff between these two factors.

\subsection{User Encoder}

For the user encoder, we employ the off-the-shelf sequence modeling tool, \ie, the self-attention mechanisms, to capture the correlations between different news historically clicked by the user. This can be formally formulated as:
\begin{align}
	\mathbf{\hat e}^*_t = \sum_\tau \beta_{t, \tau} v_u(\mathbf{e}^*_\tau) , \text{ where } \beta_{t\tau} = \frac{ \exp{ (q_u( \mathbf{e}^*_t )^T k_u( \mathbf{e}^*_\tau ) } ) }{ \sum_k \exp{ (q_u( \mathbf{e}^*_t )^T k_u( \mathbf{e}^*_k )) } },
\end{align}
where $q_u$, $k_u$, and $v_u$ denote linear transformations. In practice, we take multi-head self-attention for better performance and concatenate the outputs of multiple heads. Similarly, We obtain the final user representation $u$ by aggregating all enhanced item representations with additive attention weights:
\begin{align}
	\beta^a_t &= \frac{\tanh (k_{u,a} (\mathbf{\hat e}^*_t))}{\sum_k \tanh (k_{u,a} (\mathbf{\hat e}^*_k))}, \\
	\mathbf{u} &= \sum_t q_{u,a}( \beta^a_t ) * \mathbf{\hat e}^*_t .
\end{align}

\subsection{Training and Discussion}

Given one candidate news $c$ which we should predict how likely an user $u$ will click it, we first transform them into dense vectors $\mathbf{c}$ and $\mathbf{u}$ using IMRec, and treat $\hat{y}_{u,c} = \sigma(\mathbf{c}^T \mathbf{u})$ as the indicator and ${y}_{u,c}$ as the expected output. 
Motivated by Wu \etal, \cite{Wu_Wu_Ge_Qi_Huang_Xie_2019} and Wang \etal, \cite{Wang_Wu_Liu_Xie_2020}, we use negative sampling techniques and cross entropy loss for model training: 
\begin{align}
\mathcal{L}_{C E}=-\sum_{i=1}^{P}\log\frac{\exp(\hat{y}_{u_i,c_i}^{+})}{\exp(\hat{y}_{u_i,c_i}^{+})+\sum_{k=1}^{K}\exp(\hat{y}_{u_i,c_{i,k}}^{-})}
\end{align}
where $P$ is the number of positive training samples, $K$ is the number of negative training samples for each positive sample,
and $c_{i,k}$ means the $k$-th negative sample in the same
group with the $i$-th positive sample.

\subsection{IMRec Applied to FIM}
Noteworthy, the local impression modeling and global impression modeling modules can be \textbf{model-agnostic} and applied to any other 
CTR prediction model with ease. 
In the experiment
, we extend another SOTA method, \ie, FIM \cite{Wang_Wu_Liu_Xie_2020}, a non-sequence model that employs dilated CNN and computes the matching between each historically interacted news and the target news in a fine-grained level, to the impression-aware version, \ie, \textbf{FIM-IM}. 
Thereby, we can demonstrate the plug-and-play capability of the proposed modules. 
Specifically, in the FIM-IM model, 
only the memory network schema in the local impression module is applied to the initial word embeddings due to the high computation cost of FIM. For global impression modeling, we linearly transform global impression features into low-dimensional representations, based on which we directly compute the matching scores of each historical interacted news and the candidate news. Such matching scores are concatenated with the last layer output before the prediction. 
Given the integrated matching vector $\mathbf{s}_{u,c}$ of a user $u$ and candidate news $c$ pair, and the corresponding global impression representation $\mathbf{o}_c^*$ of the news $c$, we can calculate the final click probability as follows:
\begin{align}
\hat{y}_{u,c} = Q_2^T(\mathbf{s}_{u,c}\oplus(Q_1^T \mathbf{o}_c^* +b_1))+b_2
\end{align}
where $Q_1$, $Q_2$, $b_1$ and $b_2$ are learnable parameters, $\oplus$ means concat operation. The loss function is consistent with the NRMS-IM model.

\section{Datasets}

\subsection{Dataset Construction}
\vspace{-0.4cm}
\begin{figure}[!htbp] \begin{center}
    \includegraphics[width=\columnwidth]{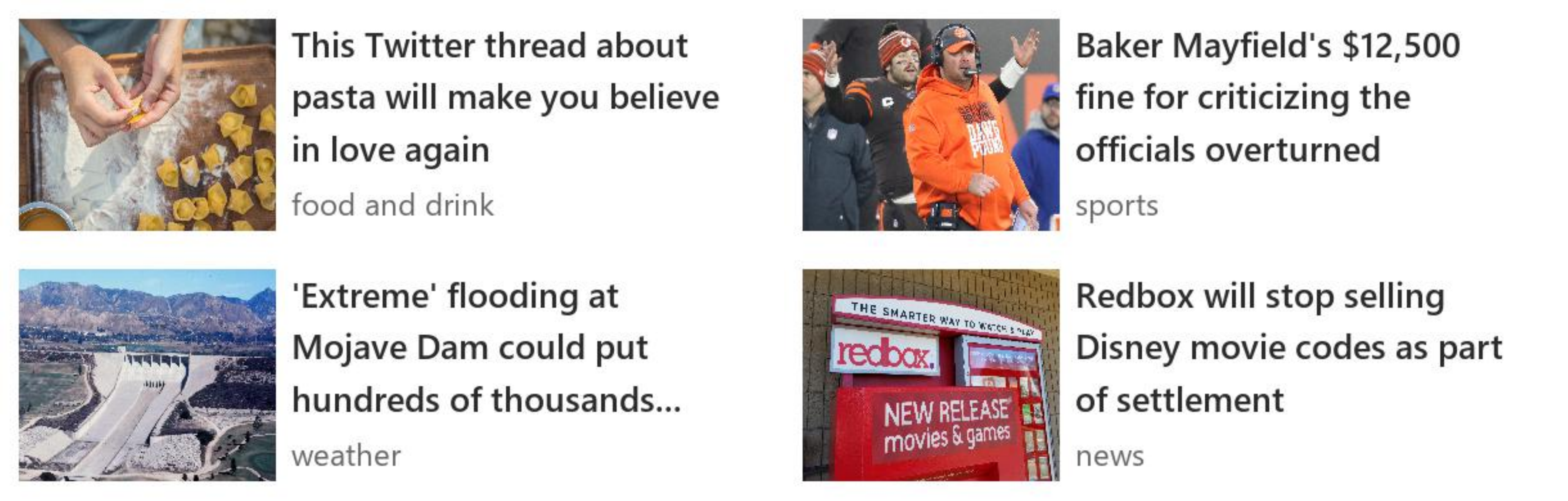}
    \caption{
    Sampled cases of news visual impression.
    	}
\label{fig:news_card}
\end{center} \end{figure}
\vspace{-0.4cm}
\noindent To the best of our knowledge, there is no news recommendation dataset suitable for impression-aware news recommendation. Therefore, we construct two benchmark datasets\footnote{The datasets will be released at \url{https://github.com/JiahaoXun/IMRec}} based on the MIND-News dataset \cite{Wu_Qiao_Chen_Wu_Qi_Lian_Liu_Xie_Gao_Wu_et_2020} automatically, 
following the styles, sizes, and spatial arrangement of different fields according to the visual impression presented in \cite{Wu_Qiao_Chen_Wu_Qi_Lian_Liu_Xie_Gao_Wu_et_2020} and the HTML code of the Microsoft news platform.

To extract news impressions, we have crawled the cover image from the given news URL and then combined news images with texts (title and category) to generate news visual impressions, as shown in figure \ref{fig:news_card}. The size of our news cards is 615*195px with white background. All news images were resized to 200*165px and were pasted at the location (15, 15) to (215, 180). News title starts at the location (215, 180) of background with 10.5px line spacing. The title lines of each news are no more than 3. Words in each line are no more than 27 characters. Font and size of title words are \texttt{seguisb} and 27, respectively. The news category starts at the location (227.725, 142.5) of background. Font and size of the category are segoeui and 24, respectively. Especially if the news does not have an image or its URL is unavailable, the image area would be empty. Since the MIND-news dataset contains large and small versions, we generate visual impressions and construct the IM-MIND-Large and IM-MIND-Small accordingly.

\subsection{Dataset Statistics}
\begin{table}[!t]
	\centering  
	\caption{Statistics of IM-MIND-Small and IM-MIND-Large.}  
	\begin{tabular}{l c l c}  
		\toprule  
		& & & \\[-10pt]
		Dataset  & Small & Users & 94057\\
		& & & \\[-10pt]  
		News& 65238 & News$_{\rm \,image}$ & 27244 \\  
		& & & \\[-10pt]  
		Avg. clicked news & 21.66 &Avg. clicked News$_{\rm \,image}$ & 8.82 \\
		& & & \\[-10pt]  
        Avg. title lines & 2.76 &Avg. words per line & 4.12\\
        \midrule
        & & & \\[-10pt]
        Dataset  & Large & Users & 876956 \\
		& & & \\[-10pt]  
		News& 130379 & News$_{\rm \,image}$ & 54421 \\  
		& & & \\[-10pt]  
		Avg. clicked news & 17.03 &Avg. clicked News$_{\rm \,image}$ & 6.35 \\
		& & & \\[-10pt]  
        Avg. title lines & 2.76 &Avg. words per line & 4.11 \\
		\bottomrule
	\end{tabular}
	\label{dataset}  
\vspace{-0.4cm}
\end{table}
The detailed statistics of the IM-MIND-Small and IM-MIND-Large dataset are summarized in table \ref{dataset}. We use \textbf{{News$_{\mathbf{ image}}$}} to denote the news that contains the cover image. The whole datatset contains 876956 users and 130379 news articles. There are 54421 available news images among all the news. The number of avg./max./min./med. clicked news are 17.35, 801, 0 and 10. The number of avg./max./min./med. clicked news with the image are 6.54, 356, 0, and 4. The avg./max. lines of title are 2.76 and 3. The avg./max./min. words per title line are 4.12, 15, and 0 (punctuation only). 

\begin{table*}[h]
\centering
    \caption{Overall performance comparison with state-of-the-art news recommenders. }
{\setlength{\tabcolsep}{0.83em}\begin{tabular}{ll ccccccccccc}
\toprule

  Datasets  & Metric  & DeepFM  & DKN &  NPA& LSTUR  & NRMS & FIM & \textbf{NRMS-IM} & \textbf{FIM-IM} & Improv.   \\
    \midrule
   
  \multirow{4}{*}{MIND-Small} 
   &  AUC    &0.6542 &0.6290 &0.6465 &\underline{0.6587} &0.6585   &0.6572 &0.6619 &\textbf{0.6661} &1.12\% 	  \\
   & NDCG@5  &0.3378 &0.3099 &0.3314 &0.3395 &0.3414	  &\underline{0.3424} &0.3465  &\textbf{0.3526}	  &2.98\%  \\
   & NDCG@10 &0.4025 &0.3741 &0.3947 &0.4015 &\underline{0.4051}   &0.4044 &0.4097  	&\textbf{0.4146}  	  &2.35\%  	  \\
   &  MRR    &0.3084 &0.2837 &0.3001 &0.3078 &\underline{0.3097}   &0.3091 &0.3132  	&\textbf{0.3199}  	  &3.29\%   \\
   \midrule
   
    \multirow{4}{*}{MIND-Large} 
   &  AUC    &0.6591 &0.6715 &0.6752 &0.6801  &0.6762   &\underline{0.6845}   &0.6866  &\textbf{0.6912} &0.98\%  \\
   &  NDCG@5 &0.3446 &0.3531 &0.3581 &0.3629  &0.3575   &\underline{0.3682}  &0.3688  &\textbf{0.3725} &1.17\%   \\
   & NDCG@10 &0.4070 &0.4171 &0.4217 &0.4265   &0.4224  &\underline{0.4313} &0.4317  &\textbf{0.4364} &1.18\%   \\
   &  MRR    &0.3140 &0.3206 &0.3261 &0.3290  &0.3227   &\underline{0.3313}  &0.3305  &\textbf{0.3364} &1.54\%	  \\
    \bottomrule
\end{tabular}}
    \label{tab:comparison}
\vspace{-0.1cm}
\end{table*}
%

\section{Experiments}

We analyze IMRec framework and demonstrate its effectiveness by anwering the following research questions:

\begin{itemize}[leftmargin=*]
	\item \textbf{RQ1}: How does IMRec perform compared with existing state-of-the-art news recommender systems?
	\item \textbf{RQ2}: Does global/local impression modeling all contribute to the effectiveness of base models in a model-agnostic manner?
	\item \textbf{RQ3}: How does IMRec perform in practical news recommendation scenarios (\eg, cold-start setting, unseen users)?
	\item \textbf{RQ4}: How does IMRec improve the performance internally?

\end{itemize}

\subsection{Experimental Settings} \label{sec:expdetails}

\begin{sloppypar}
	\vpara{Implementation Details.}The word embeddings are 100-dimensional and initialized using pre-trained Glove embedding vectors \cite{Pennington_Socher_Manning_2014}. We use pretrained resnet101\footnote{\url{https://pytorch.org/vision/stable/models.html}}  \cite{he2016deep} to extract local and global visual impression features. Specifically, for the visual word impression, we remove the last two layers of resnet101 and obtain the feature map with size (512, 28, 28) with solely the title region as input. We vertically divide the feature map by the lines of titles and further horizontally divide the vertically divided feature maps to have the word feature map by the length of words. We mean-pool the word feature map to have the impression feature of each word with dimension 512. For the cover image impression, we use the same pipeline except that we equally divide the feature map into 9 regions and mean-pool each region feature map to have a region feature.
For \textbf{global impression}, we remove the last layer of resnet101 and obtain a 2048 dimensional tensor representing the global impression of the whole news card.  
 The negative sampling ratio K is set to 4. Adam \cite{Kingma_Ba_2015} is used as the optimizer, the batch size is 32, and the initial learning rate is 1e-4. 
 These hyper-parameters are applied for both NRMS-IM and FIM-IM.
\begin{itemize}[leftmargin=*]
	\item \textbf{NRMS-IM.} Self-attention networks have 3 heads, and the output dimension of each head is 50. The dimension of the additive attention query vectors is 200. The maximum length of the tokenized word sequence of news title is set to 15. At most 60 browsed news are kept to construct the user’s recently reading behaviors. 
	\item \textbf{FIM-IM.} The maximum length of the tokenized word sequence of news title is set to 30, and at most 50 browsed news are kept for representing the user’s recently reading behaviors. Other hyper-parameter settings are following the original paper \cite{Wang_Wu_Liu_Xie_2020}.
\end{itemize}

\end{sloppypar}

\vpara{Evaluation Criteria.} Following \cite{Wu_Wu_An_Huang_Huang_Xie_2019b}, we employ three widely used metrics for evaluation, \ie, AUC (Area Under the ROC Curve), NDCG (Normalized Discounted Cumulative Gain), and MRR (Mean Reciprocal Rank).

\begin{table}[!t]
\centering
    \caption{Ablation studies by selectively discarding the local impression modeling module (- L\_IM) and global impression modeling module (- G\_IM). We study both NRMS-IM and FIM-IM to reveal the modal-agnostic capability of the proposed modules.}
{\setlength{\tabcolsep}{0.45em}
\begin{tabular}{l ccc ccc}
\toprule
&\multicolumn{3}{c}{ IM-MIND-Small } &\multicolumn{3}{c}{ IM-MIND-Large  } \\
\midrule
Model
  & AUC & NDCG@5   & MRR & AUC & NDCG@5  & MRR \\
    \midrule
    NRMS-IM     &0.6619 &0.3465 &0.3132 &0.6866 &0.3688 &0.3305   	         \\
    - L\_IM   &0.6612 &0.3440 &0.3108 &0.6837 &0.3674 &0.3295   	         \\
    - G\_IM   &0.6591 &0.3427 &0.3124 &0.6809 &0.3616 &0.3257   	         \\
    NRMS      &0.6585 &0.3414 &0.3097 &0.6762 &0.3575 &0.3227             \\
        \midrule
    FIM-IM    &0.6661 &0.3526 &0.3199 &0.6912 &0.3725 &0.3364          \\
    - L\_IM   &0.6640 &0.3479 &0.3144 &0.6909 &0.3704 &0.3349         \\
    - G\_IM   &0.6629 &0.3512 &0.3183 &0.6902 &0.3689 &0.3323	    \\
    FIM       &0.6572 &0.3424 &0.3091 &0.6845 &0.3682 &0.3313   	         \\
    \bottomrule
\end{tabular}}%

    \label{tab:ablation}
\vspace{-0.4cm}
\end{table}

\vpara{Comparison Baseline Methods.} For a comprehensive comparison to NRMS-IM and FIM-IM, we incorporate state-of-the-art baseline methods concerning both manual feature-based approaches and neural recommendation ones:
\begin{itemize}[leftmargin=*]
	\item \textbf{DeepFM} \cite{Xue_Dai_Zhang_Huang_Chen_2017}. DeepFM parallelly combines deep neural network and factorization machine. We implement it using the same feature as the LibFM.
	\item \textbf{DKN} \cite{Wang_Zhang_Xie_Guo_2018}. DKN leverages entity embeddings from knowledge graphs as external knowledge for news recommendation.
	\item \textbf{NPA} \cite{Wu_Wu_An_Huang_Huang_Xie_2019b}. NPA uses user ID embeddings to weight each word/news and thus captures important features.
	\item \textbf{LSTUR} \cite{An_Wu_Wu_Zhang_Liu_Xie_2019}. LSTUR takes the topic/subtopic as input of news encoder and uses GRU to fuse interacted news and the user embedding.
	\item \textbf{NRMS} \cite{Wu_Wu_An_Huang_Huang_Xie_2019a}. NRMS uses the multi-head self-attention to encode both news and users.
	\item \textbf{FIM} \cite{Wang_Wu_Liu_Xie_2020}. FIM employs dilated CNN and computes the matching between each historically interacted news and the target news in a fine-grained level.
\end{itemize}

\begin{table*}[!t]
\centering
    \caption{Analysis on different user/news groups. IMRec framework shows consistent improvement across various scenarios.}
\begin{tabular}{ll cc cc cc cc cc cc}
\toprule
& & \multicolumn{4}{c}{ All User } & \multicolumn{4}{c}{ Seen User } & \multicolumn{4}{c}{ Unseen User } \\
\cmidrule(lr){3-6}\cmidrule(lr){7-10}\cmidrule(lr){11-14}
   & Model & AUC & N@5 & N@10 & MRR  & AUC & N@5 & N@10 & MRR  & AUC & N@5 & N@10 & MRR   \\
    \midrule
    \multirow{2}{*}{All News}    
    & NRMS-IM &0.6866 &0.3688 &0.4317 &0.3305 &0.6901 &0.3684 	&0.4317 &0.3298 &0.6630 &0.3712 &0.4317 &0.3353 	\\ 
    & NRMS	&0.6762 &0.3575 &0.4224 &0.3227 &0.6801 &0.3577 &0.4229 &0.3225 &0.6496 	&0.3557 &0.4190 &0.3243 \\ 
    &FIM-IM &0.6912 &0.3725 &0.4364 &0.3364 	 &0.6941 &0.3716 &0.4359 &0.3351 &0.6717 	 
    &0.3784 &0.4397 &0.3452  	\\ 
    &  FIM	&0.6845 &0.3682 &0.4313 &0.3312 	 &0.6877 &0.3676 &0.4312 &0.3302 &0.6625
    &0.3723 &0.4321 &0.3379 \\ 
    \midrule
    \multirow{2}{*}{News$_{\rm \, image}$}  	    
    &  NRMS-IM &0.6984 &0.4273 &0.4869 &0.3812 &0.7010 &0.4246 	&0.4880  &0.3816 &0.6794 &0.4175 &0.4785 &0.3785 \\
    &  NRMS	 &0.6909 &0.4149 &0.4798 &0.3733 &0.6939 &0.4163 	&0.4813  &0.3741 &0.6691 &0.4050 &0.4691 &0.3673 		\\ 
    &FIM-IM	&0.6991 &0.4225 &0.4866 &0.3802 	 &0.7021 &0.4240 &0.4882 &0.3811 &0.6768 	 
    &0.4117 &0.4751 &0.3740 		\\ 
    & FIM	&0.6974 &0.4218 &0.4862 &0.3815 	 &0.7005 &0.4232 &0.4876 &0.3822 &0.6742
    &0.4113 &0.4758 &0.3759 		\\
   	\midrule
    \multirow{2}{*}{News$_{\rm \,blank}$}
    & NRMS-IM &0.6724 &0.4819 &0.5379 &0.4208 &0.6764 &0.4814 &0.5380  &0.4203  &0.6443 &0.4859 &0.5374 &0.4246 \\
    & NRMS	&0.6575 &0.4688 &0.5266 &0.4098 &0.6620 &0.4690 &0.5273 &0.4100 &0.6257 &0.4676 &0.5220 &0.4085 		\\ 
    &FIM-IM	&0.6782 &0.4895 &0.5450 &0.4296 	 &0.6810 &0.4880 &0.5442 &0.4281 &0.6588 	 
    &0.4998 &0.5507 &0.4398 		\\ 
    &  FIM	&0.6697 &0.4806 &0.5378 &0.4210 	 &0.6731 &0.4797 &0.5374 &0.4201 &0.6464
    &0.4870 &0.5405 &0.4275 		\\
    \bottomrule
\end{tabular}
    \label{tab:scenario}
\vspace{-0.4cm}
\end{table*}

\vspace{-0.4cm}
\subsection{Overall Results (RQ1)} Table \ref{tab:comparison} lists the comparison results of NRMS-IM and FIM-IM with state-of-the-art neural recommendation methods on the MIND-Small and MIND-Large datasets. From the results, we can find that:
\begin{itemize}[leftmargin=*]
	\item Overall, the results across multiple evaluation metrics consistently indicate that NRMS-IM and FIM-IM \textbf{both} achieve better results against various SOTA designs. We note that these improvements are significant and comparative to the improvement of recent SOTAs (\eg, FIM over NRMS).
	\item Surprisingly, DeepFM achieves competitive performance on the MIND-Small dataset and outperforms many advanced designs like LSTUR with GRU and NPA with the attention mechanism. However, it achieves significantly inferior results on the large-scale 
	MIND-Large dataset. The reason might be that FM based methods might fail to handle highly sparse and complex correlations. In contrast, NRMS-IM and FIM-IM achieve consistently convincing results on two datasets.
	\item Compared to DKN that also exploits additional information (\ie, entities in a knowledge graph) to enhance news representation learning. FIM-IM shows a clear advantage over it on two datasets. Noteworthy, FIM-IM improves DKN by AUC +0.0371 (relatively 5.9\%), NDCG@5 +0.0427 (relatively 13.7\%), NDCG@10 +0.0405 (relatively 10.8\%) and MRR +0.0362 (relatively 12.8\%) on the MIND-Small dataset. These results show that, compared to further enhancing the semantic understanding itself like DKN, it might be more promising to introduce visual impressions that explicitly 
	get close to the user's click decision process.
	\item Compared to other attention-based approaches, \ie, NPA and NRMS, NRMS-IM also exhibits better performance, especially on the large-scale 
	MIND-Large dataset. These results basically indicate that modeling the semantic-impression correlations (memory attending) can help improve the semantic-semantic correlation modeling.
	\item FIM-IM considers a strong baseline FIM that uses CNN as building blocks and further equips it with impression modeling modules. FIM-IM achieves state-of-the-art results with substantial improvement, demonstrating that the proposed local/global impression modeling can improve a ranking baseline with arbitrary architectures in a plug-and-play manner.
\end{itemize}

\begin{table}[!t]
\centering
    \caption{Performance on NRMS-IM by varying the percentage of visual impression used in training.}
{\setlength{\tabcolsep}{0.45em}
\begin{tabular}{l cccc}
\toprule
 Percentage
   & AUC & NDCG@5  & NDCG@10  & MRR  \\
    \midrule
    100\%	&0.6866	&0.3688 &0.4317 &0.3305  	         \\
    75\%	&0.6846	&0.3673 &0.4307 &0.3297 	         \\
    50\%	&0.6832	&0.3653 &0.4287 &0.3273 	       \\
    25\%	&0.6815	&0.3652 &0.4284 &0.3275          \\
    0\%		&0.6762	&0.3575 &0.4224 &0.3227 	        \\
    \bottomrule
\end{tabular}}%

    \label{tab:percentage}
\vspace{-0.4cm}
\end{table}

\subsection{Model Analysis (RQ2, RQ3)}

\subsubsection{Analysis on key building blocks (Ablation Study).} Local impression modeling and global impression modeling are two key components of IMRec framework. We conduct the ablation study on them to reveal the efficacy of the architectures and the benefits of incorporating local/global impression information. Specifically, we selectively discard the local impression modeling module and the global impression modeling module from NRMS-IM to generate ablation architectures, \ie, - L\_IM, and - G\_IM, respectively. We also conduct another ablation study on the FIM-IM model to show the model-agnostic capability of these two modules. The results are shown in Table \ref{tab:ablation}. We can observe that:

\begin{itemize}[leftmargin=*]
	\item Removing either L\_IM or G\_IM leads to performance degradation, and removing both modules (\ie, the base model) leads to the worst performance. These results demonstrate the effectiveness of the proposed two modules as well as the benefits of introducing visual impressions for news recommendation. We attribute this superiority to the fact that we can explicitly 
	get close to the click decision-making process by modeling the interactions of visual impression and semantic understanding of news titles.
	\item Removing G\_IM leads to more performance drops than removing L\_IM. This means that directly modeling the visual cues in the impression image without capturing these cues' spatial arrangement might be inferior to visual impression modeling.
	\item The results are consistent across different baselines, which demonstrates that the proposed two modules can easily boost a recommendation model in a plug-and-play and model-agnostic manner.
\end{itemize}

\begin{figure*}[t] \begin{center}
    \includegraphics[width=0.9\textwidth]{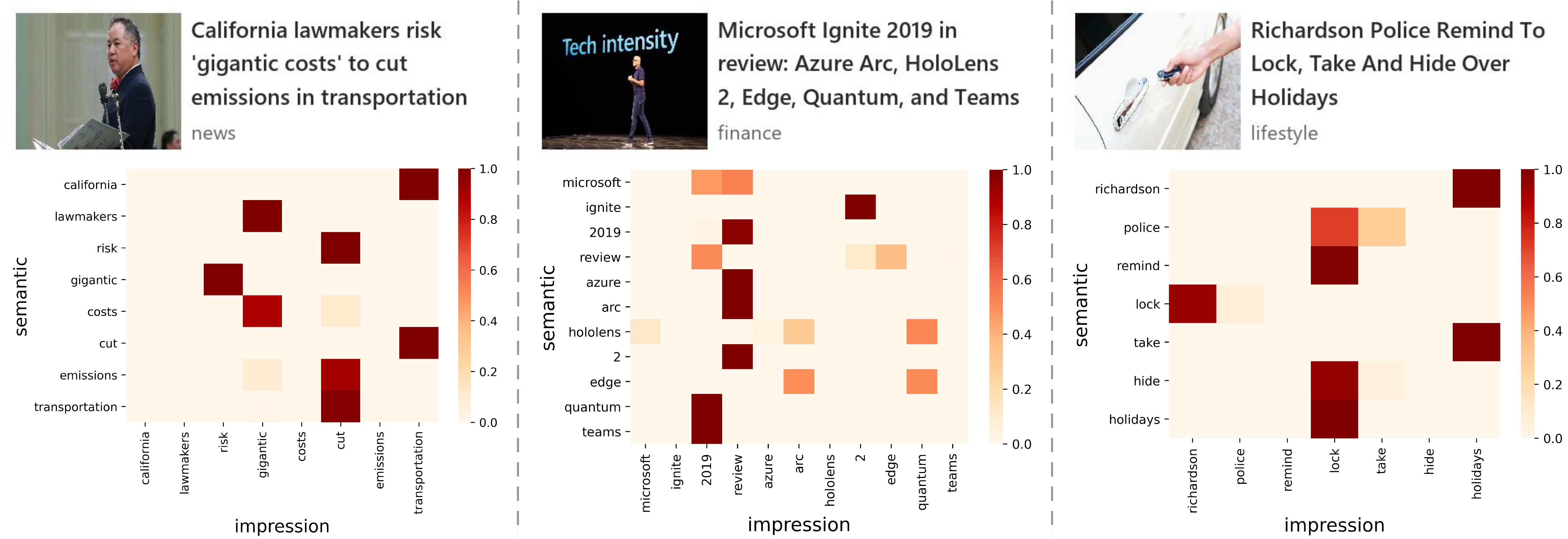}
    \caption{
    	Visualization of impression-semantic correlations by plotting the memory attending weights.
	}
\label{fig:casestudy}
\vspace{-0.5cm}
\end{center} \end{figure*} 

\subsubsection{Analysis on different user/news groups.}
In real-world news recommendation platforms, there are always unseen users beyond the training set and news without cover images. To reveal the effectiveness of impression modeling on different recommendation scenarios, we take an in-depth analysis of these two factors. The results are shown in Table \ref{tab:scenario}. For brevity,  we use \textbf{{News$_{\,\mathbf{image}}$}} to denote the news that contains the cover image and \textbf{News$_{\,\mathbf{blank}}$} to denote the news without the cover image. We can see that:
\begin{itemize}[leftmargin=*]
	\item We observe a consistent improvement of NRMS-IM/FIM-IM over the base model across various scenarios, which further demonstrates the effectiveness of IMRec framework and especially the generalization capability across different settings.
	\item The results of the two models on unseen users are worse than those on seen users, which is a reasonable result. However, we notice that the improvement of NRMS-IM over the NRMS model on \textbf{unseen} users is consistently more significant than the other user group. For example, NRMS-IM achieves 0.0155 (relatively 4.36\%) NDCG@5 improvement on unseen users (All News setting) and 0.0107 (relatively 2.99\%) NDCG@5 improvement on seen user (All News setting). Since incorporating multi-modal content is essential for cold-start setting (unseen users), these results show that the introduction of visual impression is a promising direction for news recommendation.
	\item Both two models yield better results on the news$_{\rm \, image}$  than on the news$_{\rm \, blank}$. We attribute this phenomenon to the fact that users might click the news$_{\rm \, blank}$ and eventually find that the news turns out to be less attractive. In other words, click behaviors on the news$_{\rm \, image}$ are less noisy than the news$_{\rm \, blank}$, and users' interests in the news$_{\rm \, image}$ are more consistent and easy to capture. Surprisingly, the improvement of impression modeling on the news$_{\rm \, blank}$ is more significant than the others. Considering that the interests in the news$_{\rm \, blank}$ are generally harder to capture by only using title texts, the results indicate the advantage of impression modeling in dealing with such news (\eg, the visual words might help attract users’ attention). Overall, impression modeling yields improvement on all news.
	
\end{itemize}

\subsubsection{Analysis on the percentage of visual impressions used in training.} For this experiment, we disregard the visual impressions of randomly sampled 25\%, 50\%, and 75\% news in training. In other words, news with visual impressions being masked will be represented solely by the semantic meaning of its title. We conduct the experiment on the IM-MIND-Large dataset with NRMS-IM. As shown in Table \ref{tab:percentage}, the metrics grow monotonically as the percentage of news with visual impression increases, which suggests that IMRec framework boosts the performance of the base model by effectively modeling the visual impression.

\vspace{-0.2cm}
\subsection{Qualitative Analysis (RQ4)} The above analysis quantitatively shows the effectiveness of impression-aware news recommendation. We take a further step to reveal how IMRec framework internally improves the performance of semantic-only news recommendation systems. As shown in Figure \ref{fig:casestudy}, we plot the memory attending weights of each impression word on each textual semantic word in the local impression module, which explicitly indicates the semantic-impression correlations in the fine-grained level. We note that the cases are sampled from the IM-MIND-Large 
dataset and are unseen by NRMS-IM during training. Since news with the cover image will intuitively enhance the semantic representation by providing an additional modality, we disregard them here and are more interested in the impression words, which are harder to leverage. Based on the visualization, we can find that:
\begin{itemize}[leftmargin=*]
	\item Impression words that are at the left beginning of each line (\eg, \texttt{review} in the second case, \texttt{lock} in the third case) obtain more attention weights than the others typically. This finding is intuitive, as users read the impression title in a left-to-right manner. IMRec framework 
	automatically captures such a visual correlation pattern and accordingly enhances the semantic word representation.
	\item Semantic words are more likely to attend the impression words that are spatially closer to the corresponding impression words. For example, in the third case, semantic words \texttt{Richardson} and \texttt{take} both attend to the impression word \texttt{holidays}, which are all at the beginning of lines. This result further demonstrates that IMRec framework captures the impression cues besides the sequential dependencies in semantics.
	\item A few impression words obtain the most attention, showing that IMRec framework succeeds in capturing the critical points in the impression rather than roughly attending to all impression cues.
\end{itemize}

\vspace{-0.1cm}
\section{Conclusion and Future Work} \label{sec:conclusion}
In this work, we investigate users' decision-making process when browsing and clicking news, and propose the visual impression-aware modeling framework, \ie, IMRec, for multi-modal news recommendation. IMRec explicitly gets close to the users' news reading process and simultaneously attends to local details within the impression when understanding the news title. Furthermore, IMRec fuses the multi-modal local details by considering the global arrangement of them on the impression. We contribute visual images of news impressions to MIND dataset to promote this line of research. Extensive experiments demonstrate the efficacy of IMRec in that both NRMS-IM and FIM-IM achieve better results against various state-of-the-art designs.

To the best of our knowledge, 
the work is one of the initiatives to incorporate visual impressions for news recommendation. 
By modeling the visual impressions, we might attempt to
safely disregard unnecessary modalities that are absent before users clicking the news and 
design application-specific modules for users' interest mining. 
We believe that this idea can be inspirational for other researchers
and will open up a promising direction for recommendation. We disregard more complex impression modeling designs in this paper to fairly show that introducing visual impression itself will bring many benefits. Incorporating more advanced techniques to boost performance can be a promising future work. Moreover, in many other recommendation domains, few works investigate users' click decision-making process, hence we plan to extend our idea in these domains in the future.

\section{ACKNOWLEDGEMENTS}
This work was supported in part by the National Key R\&D Program of China under Grant No.2020YFC0832505, National Natural Science Foundation of China under Grant No.61836002, No.62072397 and Zhejiang Natural Science Foundation under Grant No.LR19F020006.

\newpage
\newpage
\balance
\bibliographystyle{ACM-Reference-Format}
\bibliography{sections/9.citations}

\end{document}